# Ψ – VLASOV EQUATION

**E.E. Perepelkin[a,b,d], B.I. Sadovnikov[a], N.G. Inozemtseva[b,c], I.I. Aleksandrov[a,b]**

[a] *Faculty of Physics, Lomonosov Moscow State University, Moscow, 119991 Russia*
[b] *Moscow Technical University of Communications and Informatics, Moscow, 123423 Russia*
[c] *Dubna State University, Moscow region, Dubna,141980 Russia*
[d] *Joint Institute for Nuclear Research, Moscow region, Dubna,141980 Russia*

**Abstract**

A new equation for describing physical systems with radiation is obtained in this paper. Examples of such systems can be found in plasma physics, accelerator physics (synchrotron radiation) and astrophysics (gravitational waves). The new equation is written on the basis of the third Vlasov equation for the probability density distribution function of kinematic quantities: coordinates, velocities and accelerations. The constructed new Vlasov Ψ - equation makes it possible to describe naturally dissipation systems instead of phenomenological modifications of the second Vlasov equation, and to construct conservative difference schemes in numerical simulation.

**Introduction**

Based on the first principle – the law of conservation of probability in the generalized phase space, A.Vlasov obtained an infinite self-linking chain of equations for distribution functions $f^1(\vec{r},t)$, $f^{1,2}(\vec{r},\vec{v},t)$, $f^{1,2,3}(\vec{r},\vec{v},\dot{\vec{v}},t)$, $f^{1,2,3,4}(\vec{r},\vec{v},\dot{\vec{v}},\ddot{\vec{v}},t)$, ... of independent kinematical values $\vec{r},\vec{v},\dot{\vec{v}},\ddot{\vec{v}},...$ [1, 2]:

$$\frac{\partial f^1}{\partial t} + \text{div}_r\left[f^1\langle\vec{v}\rangle_1\right] = 0,$$

$$\frac{\partial f^{1,2}}{\partial t} + \text{div}_r\left[f^{1,2}\vec{v}\right] + \text{div}_v\left[f^{1,2}\langle\dot{\vec{v}}\rangle_{1,2}\right] = 0,$$

$$\frac{\partial f^{1,2,3}}{\partial t} + \text{div}_r\left[f^{1,2,3}\vec{v}\right] + \text{div}_v\left[f^{1,2,3}\dot{\vec{v}}\right] + \text{div}_{\dot{v}}\left[f^{1,2,3}\langle\ddot{\vec{v}}\rangle_{1,2,3}\right] = 0,$$

$$\ldots$$

$$\frac{\partial f^{1,\ldots,n}}{\partial t} + \text{div}_{\xi^1}\left[f^{1,\ldots,n}\vec{\xi}^2\right] + \text{div}_{\xi^2}\left[f^{1,\ldots,n}\vec{\xi}^3\right] + \ldots + \text{div}_{\xi^n} f^{1,\ldots,n}\langle\vec{\xi}^{n+1}\rangle_{1,\ldots,n} = 0,$$

$$\ldots$$

(i.1)

where the distribution functions are interrelated by the conditions:

$$f^0(t) = \int_{(\infty)} f^1(\vec{r},t)d^3r = \int_{(\infty)}\int_{(\infty)} f^{1,2}(\vec{r},\vec{v},t)d^3r d^3v = \int_{(\infty)}\int_{(\infty)}\int_{(\infty)} f^{1,2,3}(\vec{\xi}^{1,2,3},t)d^3\xi^1 d^3\xi^2 d^3\xi^3 =$$

$$= \int_{(\infty)}\int_{(\infty)}\int_{(\infty)}\int_{(\infty)} f^{1,2,3,4}(\vec{\xi}^{1,2,3,4},t)d^3\xi^1 d^3\xi^2 d^3\xi^3 d^3\xi^4 = \ldots,$$

(i.2)

Average kinematical values $\langle\vec{v}\rangle_1$, $\langle\dot{\vec{v}}\rangle_{1,2}$, $\langle\ddot{\vec{v}}\rangle_{1,2,3}$,... are determined by the ratios:

$$f^1(\vec{r},t)\langle\vec{v}\rangle_1(\vec{r},t) = \int_{(\infty)} f^{1,2}(\vec{r},\vec{v},t)\vec{v}d^3v,$$

$$f^{1,2}(\vec{r},\vec{v},t)\langle\dot{\vec{v}}\rangle_{1,2}(\vec{r},\vec{v},t) = \int_{(\infty)} f^{1,2,3}(\vec{r},\vec{v},\dot{\vec{v}},t)\dot{\vec{v}}d^3\dot{v}, \quad (i.3)$$

$$f^1(\vec{r},t)\langle\dot{\vec{v}}\rangle_1(\vec{r},t) = \int_{(\infty)} f^{1,2}(\vec{r},\vec{v},t)\langle\dot{\vec{v}}\rangle_{1,2}(\vec{r},\vec{v},t)d^3v,$$

$$f^{1,2,3}\langle\vec{\xi}^4\rangle_{1,2,3} = \int_{(\infty)} f^{1,2,3,4}(\vec{\xi}^{1,2,3,4},t)\vec{\xi}^4 d^3\xi^4,$$

...

The hierarchical structure of chain (i.1) is determined by different levels of completeness of information about the kinematics of the system. For example, the first equation (i.1), which is also known as the equation of continuity, defines the probability (mass, charge) density distribution function $f^1(\vec{r},t)$ in a coordinate space.

The first equation is used in continuum mechanics, field theory and electrodynamics. The second equation for function $f^{1,2}(\vec{r},\vec{v},t)$ is known as the Vlasov equation for a self-consistent field and finds wide application in statistical physics [3-5], solid-state physics [6-8], plasma physics [9-11], accelerator physics [12-14] and astrophysics [15-20]. Unlike the first equation, the second equation describes a physical system in a wider space – a phase space. When considering quantum systems in the phase space, the second Vlasov equation using the Vlasov-Moyal approximation [21] transforms into the known Moyal equation [22] for the Wigner function [23, 24].

Finding the distribution functions $f^1$, $f^{1,2}$, $f^{1,2,3}$, $f^{1,...,n}$ included in (i.1) requires the chain to be cut off on some equation and a dynamic approximation for kinematical averages $\langle\vec{v}\rangle_1$, $\langle\dot{\vec{v}}\rangle_{1,2}$, $\langle\ddot{\vec{v}}\rangle_{1,2,3}$,... to be introduced.

The chain cut-off on the first equation and representation of the vector field $\langle\vec{v}\rangle_1$ according to the Helmholtz theorem as a superposition of the vortex $\vec{A}$ and potential $\nabla_r\Phi$ fields:

$$\langle\vec{v}\rangle_1(\vec{r},t) = -\alpha\nabla_r\Phi(\vec{r},t) + \gamma\vec{A}(\vec{r},t), \quad (i.2)$$

leads to the Hamilton-Jacobi equation [25]:

$$-\hbar\frac{\partial\varphi}{\partial t} = \frac{m}{2}|\langle\vec{v}\rangle_1|^2 + e\chi = H, \quad (i.3)$$

$$e\chi \stackrel{det}{=} U + Q + \frac{e^2}{2m}|\vec{A}|^2, \qquad Q = \frac{\alpha}{\beta}\frac{\Delta|\Psi|}{|\Psi|} = -\frac{\hbar^2}{2m}\frac{\Delta|\Psi|}{|\Psi|},$$

where $\alpha \stackrel{det}{=} -\frac{\hbar}{2m}$, $\beta \stackrel{det}{=} \frac{1}{\hbar}$, $\gamma \stackrel{det}{=} -\frac{e}{m}$. Value Q is a quantum potential, which is used in the de Broglie-Bohm pilot wave theory [26-28]. Scalar potential $\Phi = 2\varphi + 2\pi k$, $k \in \mathbb{Z}$ in which function $\varphi$ is the phase of wave function $\Psi(\vec{r},t) = \sqrt{f_1}e^{i\varphi}$. Wave function $\Psi(\vec{r},t)$ satisfies the Schrödinger equation for the scalar particle in the electromagnetic field [25, 29]:

$$\frac{i}{\beta}\frac{\partial \Psi}{\partial t} = -\alpha\beta\left(\hat{p} - \frac{\gamma}{2\alpha\beta}\vec{A}\right)^2 \Psi + V\Psi, \qquad (i.4)$$

$$V \stackrel{\text{det}}{=} \frac{1}{2\alpha\beta}\frac{|\gamma\vec{A}|^2}{2} + U, \quad \hat{p} \stackrel{\text{det}}{=} -\frac{i}{\beta}\nabla, \quad \hat{p}^2 = -\frac{1}{\beta^2}\Delta,$$

with the $\vec{B} = \text{rot}_r \vec{A}$ corresponding to the magnetic induction and equation (i.3) may be represented in the form [25, 29]:

$$\frac{d}{dt}\langle\vec{v}\rangle_1 = -\gamma\left(\vec{E} + \langle\vec{v}\rangle_1 \times \vec{B}\right), \ \vec{E} = -\frac{\partial}{\partial t}\vec{A} - \nabla\chi, \qquad (i.5)$$

where $\text{rot}_r \vec{E} = -\frac{\partial \vec{B}}{\partial t}$ and $\text{div}_r \vec{B} = 0$.

The equation cut-off (i.1) on the second equation with the use of the Vlasov approximation for the vector field of the acceleration flux $\langle\dot{\vec{v}}\rangle_{1,2}$ [1]

$$\langle\dot{\vec{v}}\rangle_{1,2} = \frac{1}{m}\vec{F}, \qquad (i.6)$$

where $\vec{F}$ − force – is widely known in plasma physics $\vec{F} = -q\nabla_r\chi + q\langle\vec{v}\rangle_1 \times \vec{B}$ and in statistical physics $\vec{F} = -\nabla_r U$. The disadvantage of approximation (i.6) is the absence of the dependence of the right-hand side on velocity $\vec{v}$, which should be present in the general case due to definition (i.3) for value $\langle\dot{\vec{v}}\rangle_{1,2}$. The mentioned disadvantage disappears when the Vlasov-Moyal approximation is used [21]:

$$\langle\dot{v}_\mu\rangle_{1,2} = \sum_{n=0}^{+\infty}\frac{(-1)^{n+1}(\hbar/2)^{2n}}{m^{2n+1}(2n+1)!}\frac{\partial^{2n+1}U}{\partial x_\mu^{2n+1}}\frac{1}{f^{1,2}}\frac{\partial^{2n} f^{1,2}}{\partial v_\mu^{2n}}, \qquad (i.7)$$

which, when averaged over the velocity space, transforms into the Vlasov approximation (i.6): $\langle\langle\dot{v}_\mu\rangle\rangle = \langle\dot{v}_\mu\rangle_1 = -\frac{1}{m}\frac{\partial U}{\partial x_\mu}$. The use of approximation (i.7) transforms the second Vlasov equation (i.1) into the Moyal equation [22] for the Wigner function $f^{1,2}(\vec{r},\vec{v},t) = mW(\vec{r},\vec{p},t)$:

$$\frac{\partial W}{\partial t} + \frac{1}{m}(\vec{p},\nabla_r)W - (\nabla_r U, \nabla_p W) = \sum_{l=1}^{+\infty}\frac{(-1)^l(\hbar/2)^{2l}}{(2l+1)!}U\left(\overleftarrow{\nabla}_r, \overrightarrow{\nabla}_p\right)^{2l+1}W. \qquad (i.8)$$

On the one hand, when using the classical approximation ($\hbar \ll 1$), the Vlasov-Moyal approximation (i.7) transforms into the Vlasov approximation (i.6), and the second Vlasov equation (i.1)/Moyal (i.8) transforms into the classical Liouville equation.

On the other hand, averaging the Vlasov-Moyal approximation (i.7) over the velocity space corresponds to the «loss» of information and to the transition from the phase space to the coordinate one, which neutralizes the quantum corrections in sum (i.7) and leads to the classical Vlasov approximation (i.6).

The third Vlasov equation (i.1) for function $f^{1,2,3}(\vec{r},\vec{v},\dot{\vec{v}},t)$ contains additional information on the distribution of kinematic accelerations $\dot{\vec{v}}$. Consideration of the third equation is of significant practical importance in the description of electromagnetic radiation. Problems of accelerator physics with account taken of synchrotron radiation, modeling of plasma stability require taking into consideration electromagnetic radiation, the power of which is proportional to $\dot{v}^2$. The known Lorentz equation should be noted here describing the accelerated motion of a charged particle:

$$\dddot{\vec{v}} = \frac{6\pi\varepsilon_0 c^3}{e^2}\left(m\dot{\vec{v}} - \vec{F}_{ext}\right), \tag{i.9}$$

where $\vec{F}_{ext}$ is an external force.

Equation (i.9), unlike equation of motion (i.5), is a third order equation. The problem is that in applied problems it is the second equation is used that corresponds to the second order equation of motion (i.5). A correct description of processes with radiation (i.9) requires the use of the third Vlasov equation containing the average kinematical value $\langle\dddot{\vec{v}}\rangle$.

*The aim of this paper is to construct the second Vlasov approximation for the kinematical value $\langle\dddot{\vec{v}}\rangle$ and, as a consequence, to obtain a new form of writing the third Vlasov equation− the Vlasov $\Psi$ -equation for systems with radiation.*

The structure of the paper is as follows. In §1, based on the results of [2, 21] the authors consider the construction of the second Vlasov approximation for the average kinematical value $\langle\dddot{\vec{v}}\rangle$. As shown in §1, value $\langle\dddot{\vec{v}}\rangle$ is proportional to $\nabla_r N$, where $N$ is the radiation power. In §2, using the second Vlasov approximation obtained in §1, the Vlasov $\Psi$ -equation for distribution function $f^{1,2,3}(\vec{r},\vec{v},\dot{\vec{v}},t)$ is constructed. An example of describing a harmonic oscillator with radiation by means of distribution function $f^{1,2,3}(\vec{r},\vec{v},\dot{\vec{v}},t)$ is analyzed in detail from the standpoint of classical physics and quantum mechanics of higher-order kinematic values [29]. The conclusion contains the main results of the paper.

**§1 Derivatives of average kinematical values**

In paper [2] a dispersion chain of the Vlasov equations is considered and the ratios are obtained for the relation of average kinematical values of different orders. For average kinematical values $\langle\dot{v}\rangle_{1,2}$, $\langle\ddot{v}\rangle_{1,2,3}$ and $\langle\dddot{v}\rangle_{1,2,3}$ the following ratios are valid:

$$\frac{\partial}{\partial v_\lambda} P^2_{\mu\lambda}(1) = f^1\left[\langle\dot{v}_\mu\rangle_1 - \pi_1\langle v_\mu\rangle_1\right], \tag{1.1}$$

$$\frac{\partial}{\partial v_\lambda} P^3_{\mu\lambda}(1,2) = f^{1,2}\left[\langle\ddot{v}_\mu\rangle_{1,2} - \pi_{1,2}\langle\dot{v}_\mu\rangle_{1,2}\right], \tag{1.2}$$

$$\frac{\partial}{\partial \dot{v}_\lambda} P^4_{\mu\lambda}(1,2,3) = f^{1,2,3}\left[\langle\dddot{v}_\mu\rangle_{1,2,3} - \pi_{1,2,3}\langle\ddot{v}_\mu\rangle_{1,2,3}\right], \tag{1.3}$$

where $P_{\mu\lambda}^2$, $P_{\mu\lambda}^3$ and $P_{\mu\lambda}^4$ are momenta of the second order from distribution functions $f^{1,2}$, $f^{1,2,3}$ and $f^{1,2,3,4}$, respectively:

$$P_{\mu\lambda}^2(1,2) \stackrel{det}{=} \int_{(\infty)} f^{1,2}\left(v_\mu - \langle v_\mu \rangle_1\right)\left(v_\lambda - \langle v_\lambda \rangle_1\right) d^3v, \qquad (1.4)$$

$$P_{\mu\lambda}^3(1,2) \stackrel{det}{=} \int_{(\infty)} f^{1,2,3}\left(\dot{v}_\mu - \langle \dot{v}_\mu \rangle_{1,2}\right)\left(\dot{v}_\lambda - \langle \dot{v}_\lambda \rangle_{1,2}\right) d^3\dot{v}, \qquad (1.5)$$

$$P_{\mu\lambda}^4(1,2,3) \stackrel{det}{=} \int_{(\infty)} f^{1,2,3,4}\left(\ddot{v}_\mu - \langle \ddot{v}_\mu \rangle_{1,2,3}\right)\left(\ddot{v}_\lambda - \langle \ddot{v}_\lambda \rangle_{1,2,3}\right) d^3\ddot{v}, \qquad (1.6)$$

and $\pi_1$, $\pi_{1,2}$ and $\pi_{1,2,3}$ are differential operators of the 1st, 2nd and 3rd ranks [2]:

$$\pi_1 \stackrel{det}{=} \frac{\partial}{\partial t} + \langle \vec{v} \rangle_1 \nabla_r, \qquad (1.7)$$

$$\pi_{1,2} \stackrel{det}{=} \frac{\partial}{\partial t} + \vec{v}\nabla_r + \langle \dot{\vec{v}} \rangle_{1,2} \nabla_v, \qquad (1.8)$$

$$\pi_{1,2,3} \stackrel{det}{=} \frac{\partial}{\partial t} + \vec{v}\nabla_r + \dot{\vec{v}}\nabla_v + \langle \ddot{\vec{v}} \rangle_{1,2,3} \nabla_{\dot{v}}. \qquad (1.9)$$

Expressions (1.1)-(1.3) represent the equations of motion obtained from the second, third and fourth Vlasov equations [2]. Indeed, expressions (1.1), (1.7) imply equation of motion (i.3)/(i.5) in the hydrodynamic approximation:

$$\frac{d}{dt}\langle v_\mu \rangle_1 = \left(\frac{\partial}{\partial t} + \langle v_\lambda \rangle_1 \frac{\partial}{\partial x_\lambda}\right)\langle v_\mu \rangle_1 = -\frac{1}{f^1}\frac{\partial P_{\mu\lambda}^2}{\partial x_\lambda} + \langle \dot{v}_\mu \rangle_1, \qquad (1.10)$$

where $P_{\mu\lambda}^2$ (1.4) corresponds to the pressure tensor, and $\langle \dot{v}_\mu \rangle_1$ is determined by external force $\frac{1}{m}F_\mu$.

Expressions (1.1)-(1.3) demonstrate that the difference between the derivative from an average kinematical value of the $n$-order and an average kinematical value of the $n+1$-order is determined by momenta $P_{\mu\lambda}$. Note that momenta $P_{\mu\lambda}^2$, $P_{\mu\lambda}^3$ and $P_{\mu\lambda}^4$ are in fact covariance matrices for random kinematical values taking on values $\vec{v}$, $\dot{\vec{v}}$ and $\ddot{\vec{v}}$, respectively. If random kinematical values $v_\mu$ and $v_\lambda$ are independent then $P_{\mu\lambda}^2 = 0$ and from equation (1.1)/(1.10) it follows that

$$\frac{d}{dt}\langle v_\mu \rangle = \langle \dot{v}_\mu \rangle_1. \qquad (1.11)$$

The independence of random values is determined by the form of the distribution function. For instance [2], if distribution function $f^{n,...n+\lambda,n+1+\lambda}$, $\lambda = 0,1,...$ is even over variable $\vec{\xi}^{n+\lambda}$, i.e.

$$f^{n,\ldots,n+1+\lambda}\left(\vec{\xi}^n,\ldots,-\vec{\xi}^{n+\lambda},\vec{\xi}^{n+1+\lambda}\right) = f^{n,\ldots,n+1+\lambda}\left(\vec{\xi}^n,\ldots,\vec{\xi}^{n+\lambda},\vec{\xi}^{n+1+\lambda}\right), \tag{1.12}$$

or $P_{\alpha\beta}^{n+1+\lambda} = const$, then

$$\left\langle \xi_\alpha^{n+2+\lambda} \right\rangle_{n,\ldots,n+\lambda-1} = \pi_{n,\ldots,n+\lambda-1} \left\langle \xi_\alpha^{n+1+\lambda} \right\rangle_{n,\ldots,n+\lambda-1}. \tag{1.13}$$

The Vlasov approximation (i.6) contains two important assumptions. The first assumption is the replacement of kinematical averages in the second Vlasov equation

$$\left\langle \dot{v}_\mu \right\rangle_{1,2} = \left\langle \dot{v}_\mu \right\rangle_1. \tag{1.14}$$

The second assumption is related to the equation of motion (1.10), which reflects Newton's second law

$$\frac{d}{dt}\left\langle v_\mu \right\rangle = \frac{1}{m} F_\mu = \left\langle \dot{v}_\mu \right\rangle_1, \tag{1.15}$$

in which the transition (1.11) is made, i.e. there is no pressure force $-\frac{1}{f^1}\frac{\partial P_{\mu\lambda}^2}{\partial x_\lambda}$. The consequence of these two assumptions is the Vlasov approximation (i.6)

$$\left\langle \dot{v}_\mu \right\rangle_{1,2} = \left\langle \dot{v}_\mu \right\rangle_1 = \frac{d}{dt}\left\langle v_\mu \right\rangle = \frac{1}{m} F_\mu. \tag{1.16}$$

Such a detailed analysis of the construction of the Vlasov approximation is conducted specifically to make it clear what should be the next step and how to construct the approximation for kinematical average $\left\langle \ddot{v}_\mu \right\rangle_{1,2,3}$, which is necessary for cutting the Vlasov chain (i.1) off in the third equation.

By analogy with (1.14), the first assumption will be

$$\left\langle \ddot{v}_\mu \right\rangle_{1,2,3} = \left\langle \ddot{v}_\mu \right\rangle_{1,2}. \tag{1.17}$$

The second assumption is the absence or *weakening of correlations* between random values $\dot{v}_\mu$ and $\dot{v}_\lambda$, i.e. $P_{\mu\lambda}^3(1,2) = 0$ and equation (1.2) takes the form:

$$\left\langle \ddot{v}_\mu \right\rangle_{1,2} = \pi_{1,2} \left\langle \dot{v}_\mu \right\rangle_{1,2}. \tag{1.18}$$

For value $\left\langle \dot{v}_\mu \right\rangle_{1,2}$, there is the Vlasov-Moyal approximation (i.7), which being substituted into (1.8) gives the expression

$$\left\langle \ddot{v}_\mu \right\rangle_{1,2} = -\frac{1}{m}\pi_{1,2}\left[\frac{\partial U}{\partial x_\mu}\right] + \sum_{n=1}^{+\infty}\frac{(-1)^{n+1}(\hbar/2)^{2n}}{m^{2n+1}(2n+1)!}\pi_{1,2}\left[\frac{\partial^{2n+1}U}{\partial x_\mu^{2n+1}}\frac{1}{f^{1,2}}\frac{\partial^{2n}f^{1,2}}{\partial v_\mu^{2n}}\right]. \tag{1.19}$$

The third assumption will be a transition to the classical limit (at $\hbar \ll 1$) in expression (1.19):

$$\langle \ddot{v}_\mu \rangle_{1,2} = -\frac{1}{m}\frac{\partial}{\partial x_\mu}\left(\frac{\partial U}{\partial t} + v_\lambda \frac{\partial U}{\partial x_\lambda} + \langle \dot{v}_\lambda \rangle_{1,2} \frac{\partial U}{\partial v_\lambda}\right) = -\frac{1}{m}\frac{\partial}{\partial x_\mu}\left(\frac{\partial U}{\partial t} + v_\lambda \frac{\partial U}{\partial x_\lambda}\right),$$

$$\langle \ddot{v}_\mu \rangle_{1,2} = -\frac{1}{m}\frac{\partial N_{1,2}}{\partial x_\mu}, \quad N_{1,2} \stackrel{\text{det}}{=} \frac{\partial U}{\partial t} + v_\lambda \frac{\partial U}{\partial x_\lambda}, \quad (1.20)$$

where $N_{1,2}$ is the «radiation» power. Taking into account expressions (1.17) and (1.20), we obtain the final expression for the second Vlasov approximation

$$\langle \ddot{v}_\mu \rangle_{1,2,3} = -\frac{1}{m}\frac{\partial N_{1,2}}{\partial x_\mu}. \quad (1.21)$$

By analogy with the first Vlasov approximation (i.6), one can expand force field $-\nabla_r U$ by adding a vortex component. For instance, for the electromagnetic field (i.5) choose $-m\gamma\left(\vec{E} + \langle \vec{v} \rangle_1 \times \vec{B}\right)$ as a force.

## §2 Vlasov $\Psi$-equation

Let us write the third Vlasov equation (i.1) taking into consideration approximation (1.21), we obtain

$$\frac{\partial}{\partial t} f^{1,2,3} + v_\lambda \frac{\partial}{\partial x_\lambda} f^{1,2,3} + \dot{v}_\lambda \frac{\partial}{\partial v_\lambda} f^{1,2,3} + \langle \ddot{v}_\lambda \rangle_{1,2} \frac{\partial}{\partial \dot{v}_\lambda} f^{1,2,3} = 0,$$

$$\frac{\partial f^{1,2,3}}{\partial t} + v_\lambda \frac{\partial f^{1,2,3}}{\partial x_\lambda} + \dot{v}_\lambda \frac{\partial f^{1,2,3}}{\partial v_\lambda} - \frac{1}{m}\frac{\partial N_{1,2}}{\partial x_\lambda}\frac{\partial f^{1,2,3}}{\partial \dot{v}_\lambda} = 0, \quad (2.1)$$

where it is taken into account that $\text{div}_{\dot{v}}\langle \ddot{\vec{v}} \rangle_{1,2} = 0$. Note that integrating equation (2.1) over acceleration space $\int_{(\infty)} d^3\dot{v}$ leads to the second Vlasov equation.

Let us rewrite (assuming that the series is integrable) the Vlasov-Moyal approximation equation (i.7) as follows

$$f^{1,2}\langle \dot{v}_\mu \rangle_{1,2} = \int_{(\infty)} \dot{v}_\mu f^{1,2,3} d^3\dot{v} = \int_{(\infty)} \sum_{n=0}^{+\infty} \frac{(-1)^{n+1}(\hbar/2)^{2n}}{m^{2n+1}(2n+1)!} \frac{\partial^{2n+1}U}{\partial x_\mu^{2n+1}} \frac{\partial^{2n} f^{1,2,3}}{\partial v_\mu^{2n}} d^3\dot{v}, \quad (2.2)$$

If the following condition is met for (2.2)

$$\dot{v}_\mu f^{1,2,3} = \sum_{n=0}^{+\infty} \frac{(-1)^{n+1}(\hbar/2)^{2n}}{m^{2n+1}(2n+1)!} \frac{\partial^{2n+1}U}{\partial x_\mu^{2n+1}} \frac{\partial^{2n} f^{1,2,3}}{\partial v_\mu^{2n}}, \quad (2.3)$$

then summand $\dot{v}_\lambda \dfrac{\partial f^{1,2,3}}{\partial v_\lambda}$ in equation (2.1) can be represented as follows

$$\dot{v}_\lambda \frac{\partial f^{1,2,3}}{\partial v_\lambda} = \sum_{n=0}^{+\infty} \frac{(-1)^{n+1} (\hbar/2)^{2n}}{m^{2n+1} (2n+1)!} \frac{\partial^{2n+1} U}{\partial x_\lambda^{2n+1}} \frac{\partial^{2n+1} f^{1,2,3}}{\partial v_\lambda^{2n+1}} \qquad (2.4)$$

Taking (2.4) into account the equation (2.1) will take the following form:

$$\frac{\partial f^{1,2,3}}{\partial t} + v_\lambda \frac{\partial f^{1,2,3}}{\partial x_\lambda} + \sum_{n=0}^{+\infty} \frac{(-1)^{n+1} (\hbar/2)^{2n}}{m^{2n+1} (2n+1)!} \frac{\partial^{2n+1} U}{\partial x_\lambda^{2n+1}} \frac{\partial^{2n+1} f^{1,2,3}}{\partial v_\lambda^{2n+1}} - \frac{1}{m} \frac{\partial N_{1,2}}{\partial x_\lambda} \frac{\partial f^{1,2,3}}{\partial \dot{v}_\lambda} = 0. \qquad (2.5)$$

It should be noted that integration of (2.5) over the acceleration space $\int_{(\infty)} d^3\dot{v}$ transform it into the Moyal equation (i.8) for the Wigner function. In the classical limit ($\hbar \ll 1$), the equation (2.5) can be simplified:

$$\frac{\partial f^{1,2,3}}{\partial t} + v_\lambda \frac{\partial f^{1,2,3}}{\partial x_\lambda} - \frac{1}{m} \frac{\partial U}{\partial x_\lambda} \frac{\partial f^{1,2,3}}{\partial v_\lambda} - \frac{1}{m} \frac{\partial N_{1,2}}{\partial x_\lambda} \frac{\partial f^{1,2,3}}{\partial \dot{v}_\lambda} = 0.$$

Let us consider the simplest example of a physical system with radiation – a harmonic oscillator with stationary potential $U(\vec{r}) = \frac{m\omega^2 r^2}{2}$. Substituting potential $U(\vec{r})$ into the expression for power (1.20) and further into equation (2.1), we obtain (1.20)

$$N_{1,2} = m\omega^2 \vec{r} \cdot \vec{v}, \qquad (2.6)$$

$$\frac{\partial}{\partial t} f^{1,2,3} + \vec{v} \cdot \nabla_r f^{1,2,3} + \dot{\vec{v}} \cdot \nabla_v f^{1,2,3} - \omega^2 \vec{v} \cdot \nabla_{\dot{v}} f^{1,2,3} = 0. \qquad (2.7)$$

The solution to equation (2.7) can be found using the method of characteristics:

$$\omega^2 \vec{v} d\vec{v} = -\dot{\vec{v}} d\dot{\vec{v}} \Rightarrow \zeta(v, \dot{v}) = \omega^2 v^2 + \dot{v}^2, \qquad (2.8)$$

$$\omega^2 d\vec{r} = -d\dot{\vec{v}} \Rightarrow \vec{\eta}(x, \dot{v}) = \omega^2 \vec{r} + \dot{\vec{v}}. \qquad (2.9)$$

Thus, the solution of equation (2.7) may be represented in the form:

$$f^{1,2,3}(\vec{r}, \vec{v}, \dot{\vec{v}}) = G(\omega^2 v^2 + \dot{v}^2, \omega^2 \vec{r} + \dot{\vec{v}}), \qquad (2.10)$$

where $G = G(\zeta, \vec{\eta})$ is some function defined from the boundary conditions.

Let us consider the integrals of motion corresponding to characteristics (2.8) and (2.9). The integral of motion for characteristic (2.9) corresponds to Newton's second law:

$$m\dot{\vec{v}} = -m\omega^2 \vec{r} = -\nabla_r U. \qquad (2.11)$$

The law of conservation of kinematic energy corresponds to characteristic (2.8):

$$\frac{mv^2}{2} + \frac{m\dot{v}^2}{2\omega^2} = \frac{mv^2}{2} + \frac{m(\tau\dot{v})^2}{2} = Const, \qquad (2.12)$$

where $\omega\tau \stackrel{det}{=} 1$. The first summand in expression (2.12) corresponds to kinematic energy, and the second summand is related to energy of radiation $E_{rad}$.

Let us find out the physical nature of time $\tau$. We consider the particular solution of equation (2.11) corresponding to the motion along a closed circle with fixed radius $|\vec{r}| = r_0 = const$ with constant velocity $|\vec{v}| = v_0 = const$. The following ratios are valid:

$$U(\vec{r}) = \frac{m\omega^2 r_0^2}{2} = const, \ v = \omega r_0 = const, \ \dot{v} = \omega^2 r_0 = const. \tag{2.13}$$

Taking into consideration expressions (2.13), the conservation law (2.12) takes the form:

$$\frac{mv^2}{2} + \frac{m(\tau\dot{v})^2}{2} + U(r) = \frac{m\omega^2 r_0^2}{2} + \frac{m\omega^2 r_0^2}{2} + \frac{m\omega^2 r_0^2}{2} = \frac{3}{2}m\omega^2 r_0^2 = Const. \tag{2.14}$$

From expression (2.14) several important conclusions follow:

1. The total energy is conserved, that is, mechanical energy $E_{mech} = \frac{mv^2}{2} + U(r)$ plus radiation energy $\frac{m(\tau\dot{v})^2}{2}$ in the process of all time.
2. All three energies (kinetic, potential and radiation energy) are equal. It is especially important that the radiation energy is of the same order as the mechanical energy.

The first conclusion looks unusual, since the system must lose energy due to radiation and, as a result, the total energy must decrease. In electrodynamics, when considering radiation, the approximation is used that the total energy is much greater than the radiation energy $E_{mech} \gg E_{rad}$. Such approximation allows one to consider that the particle self-radiation does not impact its trajectory. From this, one can determine the characteristic minimum time $\tau_0$ of response to radiation. To find it $\tau_0$, let us use the Larmor formula for radiation intensity of accelerated moving particle with charge $q$ [30]:

$$I = \frac{2}{3}\frac{q^2}{4\pi\varepsilon_0 c^3}\dot{v}^2, \tag{2.15}$$

where $\varepsilon_0$ is a dielectric constant, $c$ is the speed of light. Taking into account (2.13) and (2.15), let us calculate the quantity of energy $E_{rad}$, which is radiated over time $\tau$:

$$E_{rad} = \int_0^\tau I dt = \frac{2}{3}\frac{q^2}{4\pi\varepsilon_0 c^3}\int_0^\tau \omega^4 r_0^2 dt = \frac{2}{3}\frac{q^2 r_0^2 \omega^4 \tau}{4\pi\varepsilon_0 c^3}, \tag{2.16}$$

as a result,

$$E_{mech} = m\omega^2 r_0^2 \gg E_{rad} = \frac{2}{3}\frac{q^2 r_0^2 \omega^4 \tau}{4\pi\varepsilon_0 c^3},$$

$$\tau \gg \frac{2}{3}\frac{q^2}{4\pi\varepsilon_0 mc^3} \stackrel{det}{=} \tau_0, \tag{2.17}$$

where value $\tau_0$ (2.17) is sufficiently small, for instance, for the electron the approximate value is $\tau_0 \approx 6.266 \cdot 10^{-24} s$. Over time $\tau_0$ the light manages to cover the distance $c\tau_0 = \frac{2}{3}r_e \approx 1.879 \cdot 10^{-13} cm$, where $r_e$ is the classical radius of the electron.

From ratio (2.13) it follows that time $\tau$ must be sufficiently greater than time of radiation response $\tau_0$, but in our case $E_{mech} = 2E_{rad}$, i.e. of the same order (2.14) (see Conclusion 2). Thus, ratio (2.17) is not satisfied as $\tau = 2\tau_0$. Using the Larmor formula (2.15) on such time ($\sim \tau_0$) and coordinate ($\sim r_e$) scales is an approximate, evaluative approach, since electrodynamics is no longer applicable on scales $\sim 137 r_e$ ($1/137$ is a fine-structure constant). Nonetheless, in this approximation, one more result can be obtained – the Lorentz equation.

Let us differentiate expression for the total energy (2.14) with respect to time:

$$m\vec{v} \cdot \dot{\vec{v}} + m\tau^2 \dot{\vec{v}} \cdot \ddot{\vec{v}} + \vec{v} \cdot \nabla_r U = 0. \tag{2.18}$$

In the framework of the considered approximation, we set $\tau \dot{\vec{v}} = \vec{v}$ and $\tau = \tau_0$, we obtain

$$\vec{v} \cdot \left( m\dot{\vec{v}} + m\tau_0 \ddot{\vec{v}} + \nabla_r U \right) = \vec{v} \cdot \vec{\Lambda} = 0. \tag{2.19}$$

Equation (2.19) has two solutions: $\vec{\Lambda} = \vec{0}$ and $\vec{\Lambda} \perp \vec{v}$. The condition $\vec{\Lambda} \perp \vec{v}$ can be implemented by assuming $\vec{\Lambda} = q(\vec{v} \times \vec{B})$, where $\vec{B}$ is some vector. As a result, equation (2.19) takes the form:

$$m\dot{\vec{v}} = -\nabla_r U + q(\vec{v} \times \vec{B}) - \frac{q^2}{6\pi\varepsilon_0 c^3}\ddot{\vec{v}}, \tag{2.20}$$

which coincides with the Lorentz equation (i.9).

Let us consider one-dimensional motion, for instance, along the OX axis. With such a motion, potential energy $U(x)$ changes. As a result, integral (2.12) implies that total energy $E_{rad} + E_{mech}$ is not conserved. According to integral (2.11), mechanical energy $E_{mech}$ is conserved $E_{mech} = m\omega^2 x_0^2/2$, where $x_0$ is the maximum deviation from the equilibrium position. Acceleration $\dot{v}$ changes from zero to $x_0\omega^2$. Therefore, radiation energy $E_{rad}$ varies from zero to $m\omega^2 x_0^2/2$. Thus, the ratio between energies $E_{mech}$ and $E_{rad}$ in the course of motion undergoes significant changes from $E_{mech} \gg E_{rad}$ to $E_{mech} = E_{rad}$. As a result, the question arises of the correct application of the Larmor formula and the limits of the electrodynamics description.

Despite the reached limit of applicability of electrodynamics, note that the initial statement of the problem was formulated for probability distribution function $f^{1,2,3}(\vec{r},\vec{v},\dot{\vec{v}})$, which satisfies the Vlasov $\Psi-$equation (2.7)/(2.5). The description of a system by means of a distribution function leads not only to the apparatus of statistical physics, but also to quantum mechanics. It is quantum mechanics that is a natural tool for describing a physical system at the micro-scale level. Usual quantum mechanics deals with coordinate or momentum representation. Quantum mechanics in the phase space uses the concept of quasi-density of probabilities – the Wigner function [23, 24]. Since distribution function $f^{1,2,3}(\vec{r},\vec{v},\dot{\vec{v}})$, in addition to the coordinate and velocity/momentum, depends on the acceleration, we will use quantum mechanics of higher kinematic values [29].

With transition to quantum mechanics, a question arises whether the second Vlasov approximation (1.21) is correct. On the one hand, when obtaining expression (1.21), a transition was made to the classical limit (at $\hbar \ll 1$) in expression (1.19). On the other hand, the right-hand side of expression (1.19) contains derivatives $\dfrac{\partial^{2n+1} U}{\partial x_\mu^{2n+1}}$ from the potential, which are equal to zero for a harmonic oscillator at $n>0$. The harmonic oscillator is a unique physical system [33, 21, 31, 32], for which the classical Liouville equation coincides with the quantum Moyal equation (the second Vlasov equation). Thus, the second Vlasov approximation (1.21) remains correct for the quantum harmonic oscillator as well.

Using the obtained form of solution (2.14) and the results of the paper [2, 29], let us write the expression for 1D-case of function $f^{1,2,3}(x,v,\dot{v})$:

$$f_n^{1,2,3}(x,v,\dot{v}) = \frac{(-1)^n}{2\pi\sigma_x \sigma_v} e^{-\frac{\dot{v}^2}{2\sigma_{\dot{v}}^2} - \frac{v^2}{2\sigma_v^2}} L_n\left(2\left(\frac{\dot{v}^2}{2\sigma_{\dot{v}}^2} + \frac{v^2}{2\sigma_v^2}\right)\right) \delta(\dot{v} + \omega^2 x), \tag{2.21}$$

where $n$ is the number of the quantum state; $L_n$ are the Laguerre polynomials; $\sigma_x$ and $\sigma_v$ are the standard deviations satisfying the ratios $\omega = \dfrac{\sigma_v}{\sigma_x} = \dfrac{\sigma_{\dot{v}}}{\sigma_v}$, $\sigma_x \sigma_v = \dfrac{\hbar}{2m}$. Integrating functions (2.21) over the acceleration space (i.2) gives the known Wigner function of the harmonic oscillator $f_n^{1,2}(x,v) = mW(x,p)$:

$$f_n^{1,2}(x,v) = \int_{-\infty}^{+\infty} f_n^{1,2,3}(x,v,\dot{v}) d\dot{v} = \frac{(-1)^n m}{\pi\hbar} e^{-\frac{m}{\hbar\omega}(v^2 + \omega^2 x^2)} L_n\left(\frac{2m}{\hbar\omega}(v^2 + \omega^2 x^2)\right). \tag{2.22}$$

The subsequent integration of the distribution function (2.22) over the velocity space results in probability density $f_n^1(x) = |\Psi_u|^2$, where $\Psi_n$ is a coordinate representation of the wave function of the harmonic oscillator:

$$f_n^1(x) = \int_{-\infty}^{+\infty} f_n^{1,2}(x,v) dv = \frac{1}{2^n n!} \frac{1}{\sqrt{2\pi}\sigma_x} e^{-\frac{x^2}{2\sigma_x^2}} H_n^2\left(\frac{x}{\sqrt{2}\sigma_x}\right), \tag{2.23}$$

where $H_n$ are the Hermitian polynomials. Note that ratios (i.3), (1.1)-(1.3) are satisfied in this case:

$$\langle \dot{v} \rangle_{1,2} = \langle \dot{v} \rangle_1 = -\omega^2 x,$$
$$\langle \ddot{v} \rangle_1 = \pi_1 \langle \dot{v} \rangle_1 = -\omega^2 \pi_1 x = -\omega^2 \langle v \rangle_1. \tag{2.24}$$

By analogy with distribution functions (2.21)-(2.23), the following transitions are valid for the equations from the Vlasov chain (i.1): integrating the third equation over the acceleration space gives the second equation, and integrating the second equation over the velocity space gives the first equation.

**Conclusions**

The main result of the work is the obtained modification of the third Vlasov equation – the Vlasov $\Psi$ - equation (2.1)/(2.5) for probability density distribution function $f^{1,2,3}\left(\vec{r},\vec{v},\dot{\vec{v}},t\right)$. Equation (2.1)/(2.5) allows us to take a fresh look at classical systems with radiation. Plasma and a wide range of applied problems related to thermonuclear fusion can serve as an example of such a physical system. Another field of application of equation (2.1)/(2.5) is high-energy physics, the methods of which are used to design accelerator complexes that take synchrotron radiation into account. The tasks of astrophysics associated with modeling the radiation of gravitational waves are also worth noting.

The Vlasov $\Psi$-equation may be considered as an extended version of the second Vlasov equation for the description of dissipative systems. To take into account dissipations, the second Vlasov equation is modified phenomenologically by adding summands to the right-hand side [34, 35]. In the Vlasov $\Psi$-equation, dissipation in the form of radiation is naturally contained in the equation due to approximation $\langle\ddot{\vec{v}}\rangle$.

Modeling of complex physical systems is usually performed using numerical methods. A variety of papers on the numerical solution of the Vlasov, Vlasov-Poisson and Vlasov-Maxwell equations [36-41] exist. The results obtained in this paper may find application in numerical modelling as additional conservation laws necessary for constructing conservative difference schemes. The presence of additional conservation laws is of particular importance when modelling plasma stability.

**Acknowledgements**

This research has been supported by the Interdisciplinary Scientific and Educational School of Moscow University «Photonic and Quantum Technologies. Digital Medicine».